\documentclass[
 reprint,
 twocolumn,
 amsmath,amssymb,
 aps,
floatfix,
]{revtex4-1}

\usepackage{graphicx}
\usepackage{dcolumn}
\usepackage{bm}
\usepackage{color}
\usepackage{booktabs}
\usepackage{fdsymbol}

\usepackage[margin=0.7in]{geometry}
\begin{document}

\title{Grain structure dependence of coercivity in thin films}
\author{Anton Bachleitner-Hofmann$^1$}
\author{Bernhard Bergmair$^2$}%
\author{Thomas Schrefl$^3$}
\author{Armin Satz$^4$}
\author{Dieter Suess$^1$}
\affiliation{%
$^1$ CD-Laboratory for Advanced Magnetic Sensing and Materials, Institute of Solid State Physics, TU Wien, Vienna 1040, Austria\\
$^2$ Linz Center of Mechatronics GmbH, Linz 4040, Austria \\
$^3$ Center for Integrated Sensor Systems, Danube University Krems, 2700 Wiener Neustadt, Austria\\
$^4$ Infineon Technologies Austria AG, Villach 9500, Austria
}%

\date{\today}
\setlength\heavyrulewidth{0.35ex}
\renewcommand{\arraystretch}{1.2}
\setlength\tabcolsep{1pc}

\begin{abstract}
We investigated coercive fields of 200nm x 1200nm x 5nm rectangular nanocrystalline thin films as a function of grain size D using finite elements simulations. To this end, we created granular finite element models with grain sizes ranging from 5nm to 60nm, and performed micromagnetic hysteresis calculations along the y-axis (easy direction) as well as along the x-axis (hard direction). We then used an extended Random Anisotropy model to interpret the results and to illustrate the interplay of random anisotropy and shape induced anisotropy, which is coherent on a much larger scale, in thin films.
\end{abstract}
\maketitle

\section*{Introduction}
Coercive properties of magnetic materials are widely considered to be of great importance. Particularly magnetic sensor applications often rely on thin films, where the understanding of coercive and hysteretic properties is crucial. According to the random anisotropy model, which was first introduced by Alben~\cite{alben1978random} for amorphous materials but shown by Herzer to be applicable for nanocrystalline materials as well~\cite{herzer1990grain,herzer1993nanocrystalline,herzer1995soft}, the coercive field strongly depends on the nanocrystalline grain size D. We investigated coercive fields of 200nm x 1200nm x 5nm rectangular nanocrystalline thin films as a function of D using finite elements simulations and compared the results to an extended random anisotropy model~\cite{herzer2005anisotropies}. Due to the thin film nature and the high aspect ratio in the lateral dimensions, our structure exhibits a strong shape anisotropy caused by the demagnetizing field, which also contributes to the effective anisotropy~\cite{herzer2005anisotropies,herzer2005random,suzuki2012magnetic,suzuki1998effect}.
Material parameters used in the finite elements simulations were: saturation magnetization $\mu_0M_s = 1.75T$, exchange stiffness constant $A = 1.5\times 10^{-11}\frac{J}{m}$, magnetocrystalline anisotropy constant $K_1 = 100\frac{kJ}{m^3}$ with a standard deviation of 10\% and gilbert damping parameter $\alpha = 0.1$. The high damping parameter of $\alpha = 0.1$ is justified because we were not interested in fast dynamic processes but in quasi-equilibrium hysteresis calculations with a sufficiently slow field rate of $60\frac{mT}{\mu s}$. The maximum applied external field was $\mu_0H_{j,max} = 150mT$ for both easy axis and hard axis simulations.

\section*{Micromagnetic modelling of nanocrystalline grain structure in thin films}
\begin{figure}[h]
\begin{center}
\includegraphics[width=20pc]{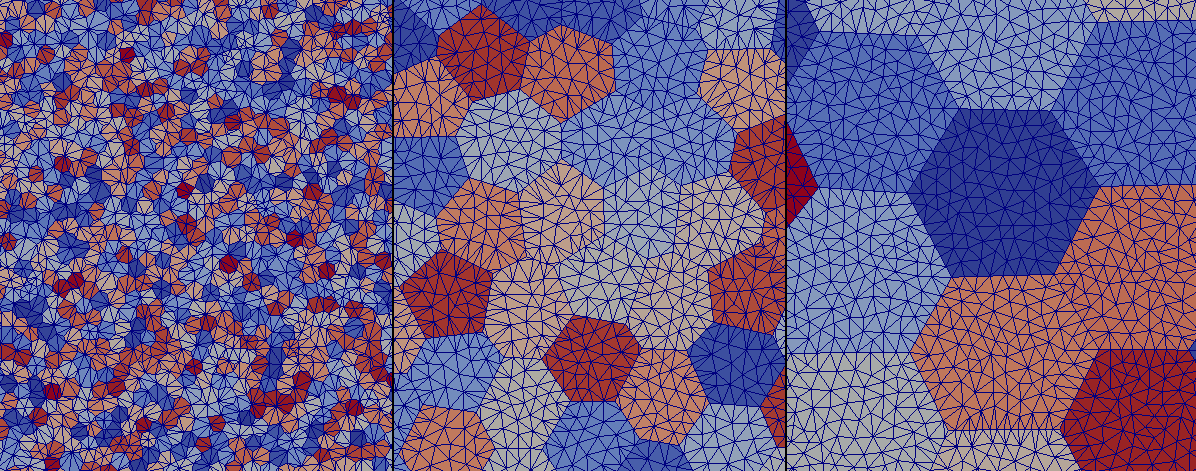}
\caption{\label{grainmodel}Visualization of our nanocrystalline finite element model for grain size distribution maximum D of 5.5nm (left), 29nm (middle) and 57.5nm (right). Grain sizes are approximately log-normal distributed with a targeted maximum D in the probability density function from 5 nm to 60 nm and a $\sigma$ of 0.4. The size of tetrahedral mesh elements is less than 5 nm for all grain sizes. The colors have no physical meaning and are meant as help to visually distinguish the grains.}
\end{center}
\end{figure}

\begin{figure}[h]
\begin{center}
\includegraphics[width=20pc]{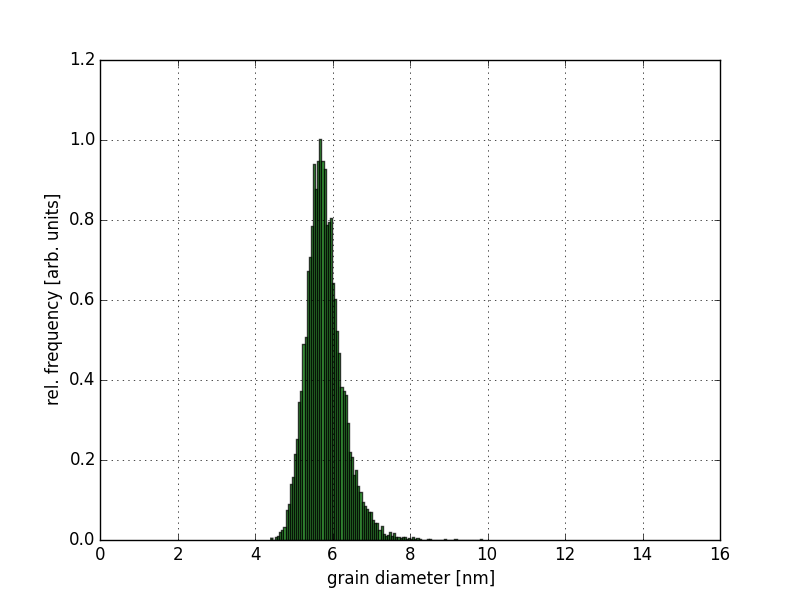}
\caption{\label{histD5} Histogram of structure with log-normal grain size distribution and a maximum in the probability density function at approximately 5.5nm. }
\vspace{1pc}
\includegraphics[width=12pc]{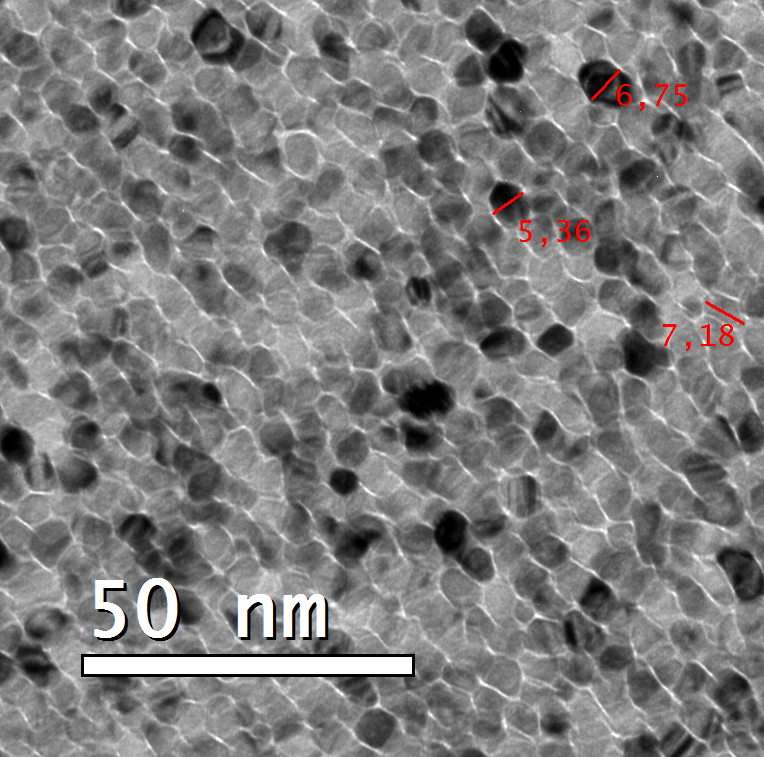}
\caption{\label{TEMimage}TEM image of CoFe thin film structure used in a GMR based sensor device. Film thickness is 5nm.}
\end{center}
\end{figure}

Grain sizes in CoFe films, as in nanocrystalline materials in general, are considered to be log-normal distributed~\cite{vopsaroiu2005preparation}. If we use micromagnetic simulations to try to predict properties based on average grain size, we also have to take that grain size distribution into account~\cite{bitoh2003random}. To this end we developed an algorithm that is able to create tetrahedral finite element meshes of thin film structures with arbitrary dimensions and arbitrary grain size distributions (Figures \ref{grainmodel}-\ref{TEMimage}) using a 2D sphere packing algorithm combined with a weighted voronoi tessellation (laguerre tessellation). The strength of the developed algorithm is that any arbitrary grain size distribution can be realized. We created structures with average grain sizes ranging from 5.5nm to 57.5nm, with a maximum tetrahedral mesh cell size of 5nm, regardless of the respective grain size. The meshes were generated using the well known TetGen package~\cite{si2015tetgen}. For small grain sizes ($<20nm$), our algorithm tends to produce grain size distributions with a probability density function maximum at slightly larger grain sizes ($\approx +0.5nm$), while for larger grain sizes the resulting distributions have their maximum at slightly lower grain size values than targeted. It should be noted that for large grain sizes, the total number of grains inside the whole particle gets very low, which leads to bad statistics and results that depend strongly on the set of random numbers used in the model creation algorithm. To interpret the grain size dependence at grains larger than the exchange length, one would therefore have to use significantly larger structures, or take the mean of a large number of simulations with the same physical parameters but different sets of random numbers used in the model creation, both of which currently exceeds our computational capacities.

\section*{Random anisotropy model}

\begin{figure}[h]
\begin{center}
\includegraphics[width=11pc]{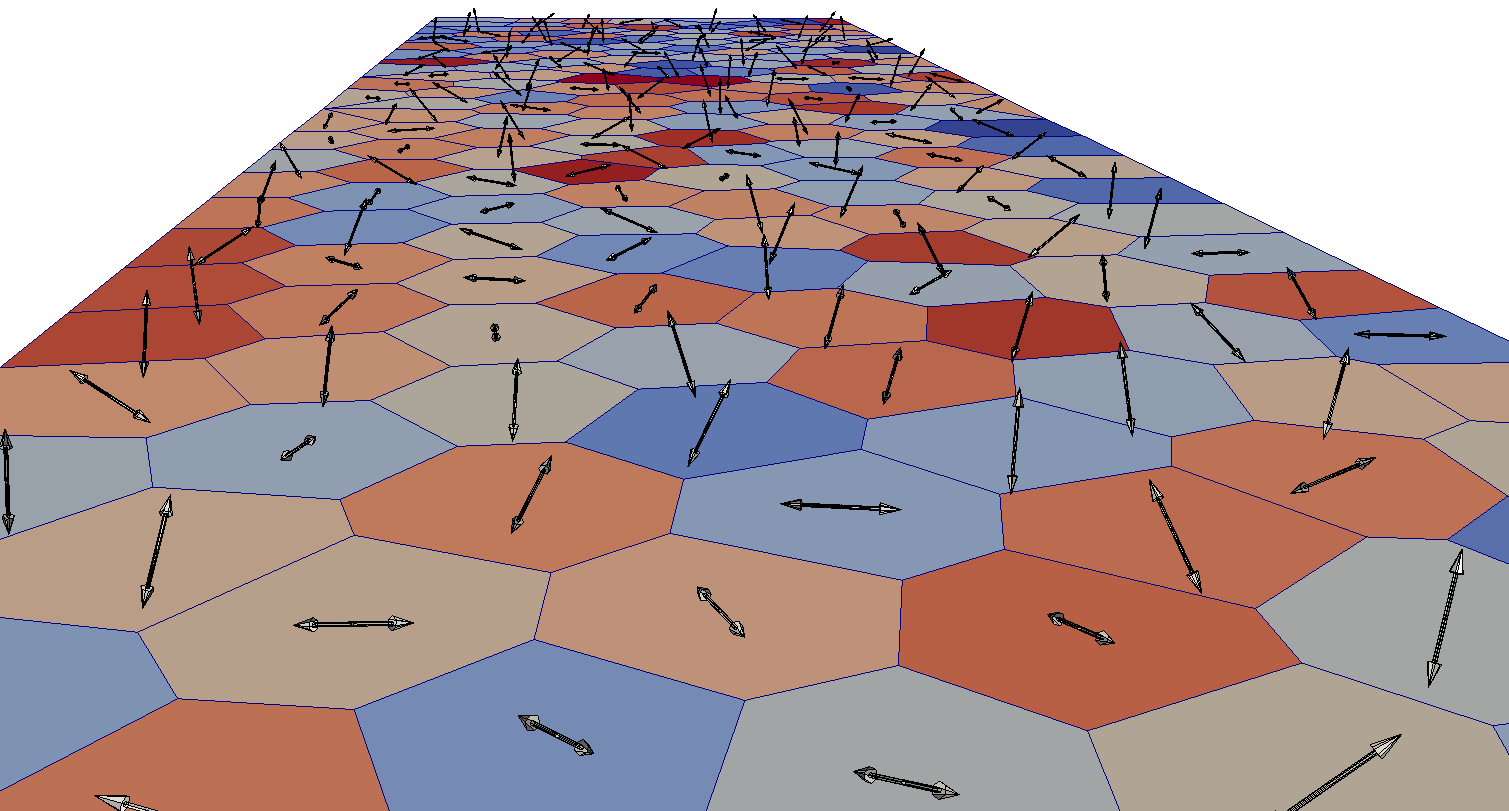}
\caption{\label{RAMgrains} Every grain has a different anisotropy axis, the orientation of which is randomly distributed in 3 dimensions. For grain diameters smaller than the exchange length, the magnetization in neighboring grains with different anisotropy axis orientation cant align with their local respective easy axes without disregarding exchange interaction intergranularly. Some common energy minimum, where the magnetization deviates from the local anisotropy axes, has to be found, leading to greatly reduced effective anisotropy.}
\vspace{1pc}
\includegraphics[width=20pc]{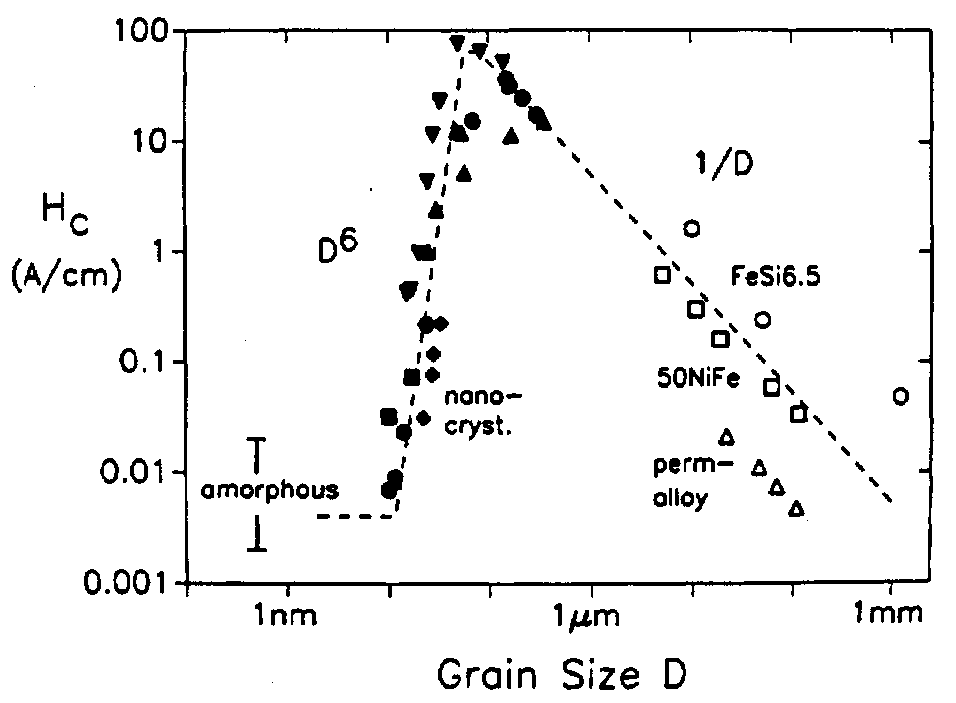}
\caption{\label{RAM-Herzer}Coercivity over grain size for various alloys ($\medblacktriangleup$: Fe-Nb-Si-B; $\medblackcircle$: Fe-Cu-Nb-Si-B; $\medblackdiamond$: Fe-Cu-V-Si-B; $\medblacksquare$: Fe-Zr-B; $\medblacktriangledown$: Fe-Co-Zr; $\medtriangleup,\medsquare$: Ni-Fe; $\medcircle$: Fe-Si), showing the $D^6$ dependence typical for bulk materials. This figure was taken from reference~\cite{herzer1993nanocrystalline} with kind permission of Giselher Herzer and IOP Publishing~\footnote{\copyright IOP Publishing.  Reproduced with permission.  All rights reserved}.}
\end{center}
\end{figure}

The well known random anisotropy model \cite{herzer1990grain} predicts coercivity as a function of nanocrystalline grain size. If the grain diameter is smaller than the intrinsic ferromagnetic exchange length $L_0$, the local crystalline anisotropies, that have different orientation in every grain (Fig. \ref{RAMgrains}), are largely suppressed because ferromagnetic exchange forces magnetic moments to align intergranularly, regardless of their local anisotropy axis, similar to the case of amorphous materials~\cite{alben1978random}. The effective magnetocrystalline anisotropy can be expressed as 
\begin{equation}
\label{eq:K}
\langle K \rangle = \langle K_1 \rangle = \frac{K_1}{\sqrt{N}}
\end{equation}
Where $\sqrt{N}$ is the number of grains within the exchange volume defined by the effective exchange length $L_{ex}$. The effective exchange length in turn depends on the effective anisotropy
\begin{equation}
\label{eq:Lex}
L_{ex} = \varphi \sqrt{\frac{A}{\langle K \rangle}}
\end{equation}
but is not to be confused with the intrinsic or natural exchange length $L_0$.
\begin{equation}
\label{eq:L0}
L_0 = \varphi \sqrt{\frac{A}{K_1}}
\end{equation}
Where $A$ is the exchange stiffness and $\varphi$ is a dimensionless proportionality factor.

\subsection*{Bulk materials}
For bulk materials, the number of grains inside the exchange volume is given by
\begin{equation}
\label{eq:Nbulk}
N=\frac{L_{ex}^3}{D^3}
\end{equation}
where $D$ is the grain diameter. If we combine \eqref{eq:K}, \eqref{eq:Nbulk} and \eqref{eq:Lex}
\begin{equation*}
\langle K \rangle = \frac{K_1}{\sqrt{N}} = K_1 \frac{D^\frac{3}{2}}{L_{ex}^\frac{3}{2}} = K_1 \frac{D^\frac{3}{2} \langle K \rangle^\frac{3}{4}}{\varphi^\frac{3}{2} A^\frac{3}{4}}
\end{equation*}
and solve for $\langle K \rangle$, we obtain the well known $D^6$ law
\begin{equation}
\langle K \rangle = \frac{K_1^4 D^6}{\varphi^6 A^3} = K_1 \left(\frac{D}{L_0}\right)^6
\end{equation}
which has been extensively studied and experimentally verified by Herzer (Fig. \ref{RAM-Herzer})\cite{herzer1990grain,herzer1993nanocrystalline,herzer1995soft,herzer2005random}.

\subsection*{Thin films}
If we apply the random anisotropy model to thin films, where the film thickness is lower than or equal to the average grain diameters, the anisotropy is averaged over 2 instead of 3 dimensions~\cite{thomas2008microstructure,herzer1991magnetization}, and thus the number of grains within one exchange volume changes to
\begin{equation}
\label{eq:Nthinfilm}
N = \frac{L_{ex}^2}{D^2}
\end{equation}
If we again combine \eqref{eq:K}, \eqref{eq:Nthinfilm} and \eqref{eq:Lex}
\begin{equation}
\label{eq:2DK1}
\langle K \rangle = \langle K_1 \rangle = \frac{K_1}{\sqrt{N}} = K_1\frac{D}{L_{ex}} = K_1 \frac{D}{\varphi}\sqrt{\frac{\langle K \rangle}{A}}
\end{equation}
we obtain the effective anisotropy for 2-dimensional systems
\begin{equation}
\label{eq:2DK1_final}
\langle K \rangle = \frac{K_1^2}{\varphi^2 A} D^2 = K_1 \left(\frac{D}{L_0}\right)^2
\end{equation}
which has been studied experimentally by Thomas et al.~\cite{thomas2008microstructure}. However they found an exponent of only 1.5 instead of 2, which they attributed to the grain size distribution~\cite{bitoh2004effect} which had not been taken into account.

\subsection*{Additional anisotropies}
\label{sec:analyticHcGs}
\begin{figure}[t]
\begin{center}
\includegraphics[width=20pc]{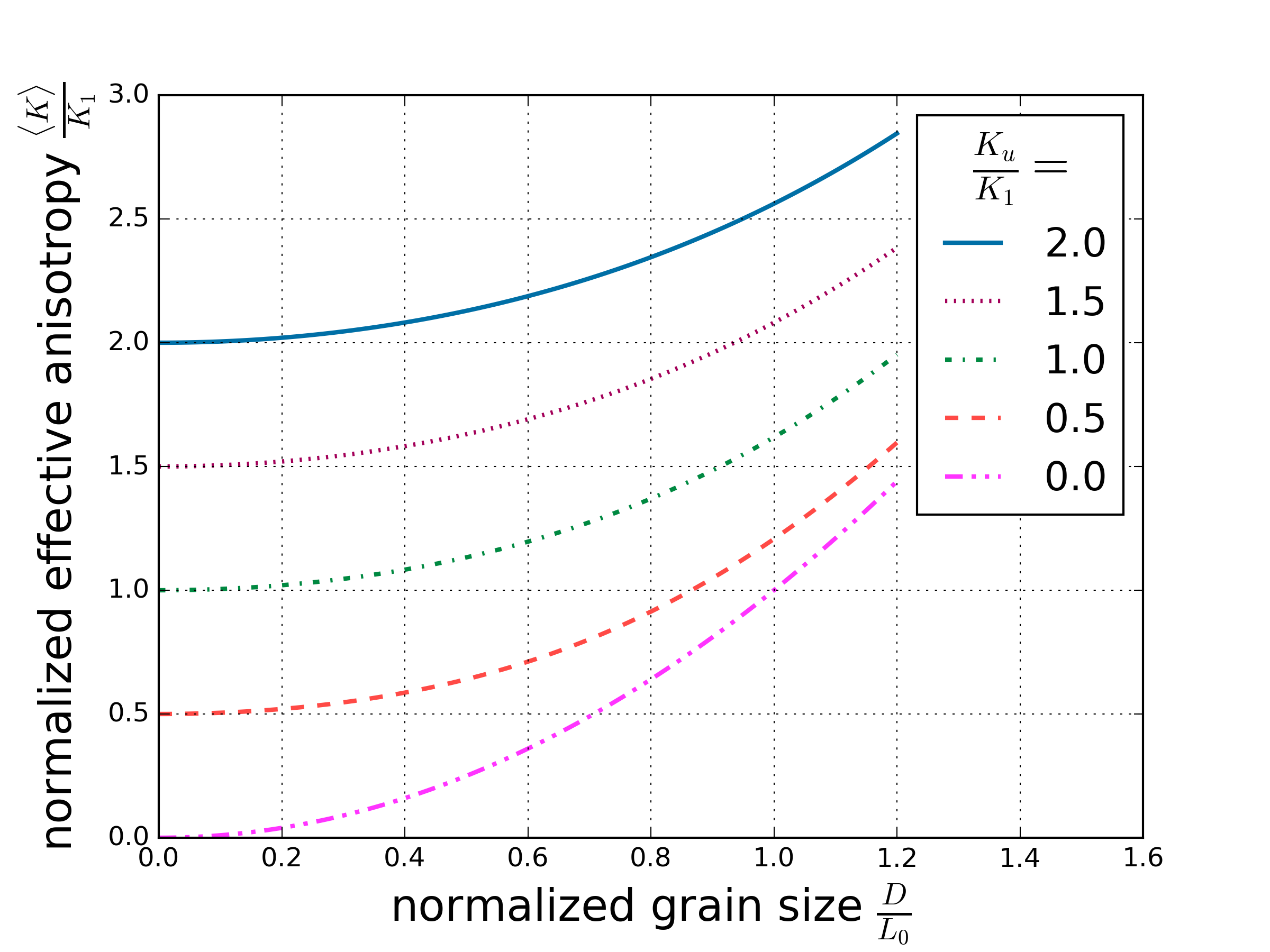}
\caption{\label{analyticKuK1}Interplay of random crystalline anisotropies $\langle K_1 \rangle$ and long-range uniform anisotropy $K_u$ in thin films~\eqref{eq:RAM2DKuModel} .}
\end{center}
\end{figure}

The structures studied in this work exhibit a high aspect ratio of $\frac{L_y}{L_x}=6$, where the magnetization is preferably oriented parallel to the long axis of the element. This shape anisotropy can thus be expressed by an effective uniaxial anisotropy with anisotropy constant $K_u$. Since the particles are large enough for the magnetization reversal to be a potentially complex multi-domain process, the anisotropy constant $K_u$ will be considerably smaller than what would be calculated by demagnetization factors ($K \propto (N_{\parallel} - N_{\perp})\frac{J_s^2}{\mu_0} \ll \frac{J_s^2}{\mu_0}$ where $\frac{J_s^2}{\mu_0}$ is used by various groups to estimate domain wall widths in thin films) and coercive fields might depend on the direction of measurement. But for the purpose of extending the Random Anisotropy Model, it is sufficient to assume uniaxial anisotropy. In that case the effective anisotropy becomes~\cite{suzuki2012magnetic,herzer2005anisotropies,suzuki1998effect}
\begin{equation}
\label{eq:K1mitKu}
\langle K \rangle \approx \sqrt{K_u^2 + \langle K_1 \rangle^2}
\end{equation}
which is based on scaling arguments and statistical considerations and was discussed by Herzer~\cite{herzer1995soft,herzer2005random} and Alben et al.~\cite{alben1978random}. Other anisotropies such as stress induced (magneto-elastic) or field induced (grain alignment) anisotropies could be incorporated analogously. \\
If we insert \eqref{eq:2DK1} in \eqref{eq:K1mitKu}, we obtain 
\begin{equation}
\langle K \rangle = \sqrt{K_u^2 + K_1^2 \frac{D^2}{\varphi^2}\frac{\langle K \rangle}{A}} = \sqrt{K_u^2 + K_1 \left(\frac{D}{L_0}\right)^2\langle K \rangle}
\end{equation}
which, for the 2-dimensional case can easily be solved for $\langle K \rangle$.
\begin{equation}
\label{eq:RAM2DKuModel}
\langle K \rangle = \frac{K_1}{2} \left(\frac{D}{L_0}\right)^2 + \sqrt{\left(\frac{K_1}{2} \left(\frac{D}{L_0}\right)^2\right)^2 + K_u^2}
\end{equation}
In the case of vanishing $K_u$, equation \eqref{eq:RAM2DKuModel} simplifies to \eqref{eq:2DK1_final}, while for vanishing grain size $D$ or vanishing $K_1$, the effective anisotropy $\langle K \rangle$ reduces to $\langle K \rangle = K_u$. The interplay of $K_1$ and $K_u$ is illustrated in Fig.~\ref{analyticKuK1}.
For the more general three dimensional case, this derivation can also be performed analytically, but leads to a lenghty and unelegant expression for $\langle K \rangle$~\cite{suzuki2008magnetic} and is usually only given for the limiting cases of negligible or dominating anisotropy $K_u$.

\subsection*{Exchange length in thin films}
\label{sec:exchangethinfilms}
\begin{figure}[t]
\begin{center}
\includegraphics[width=20pc]{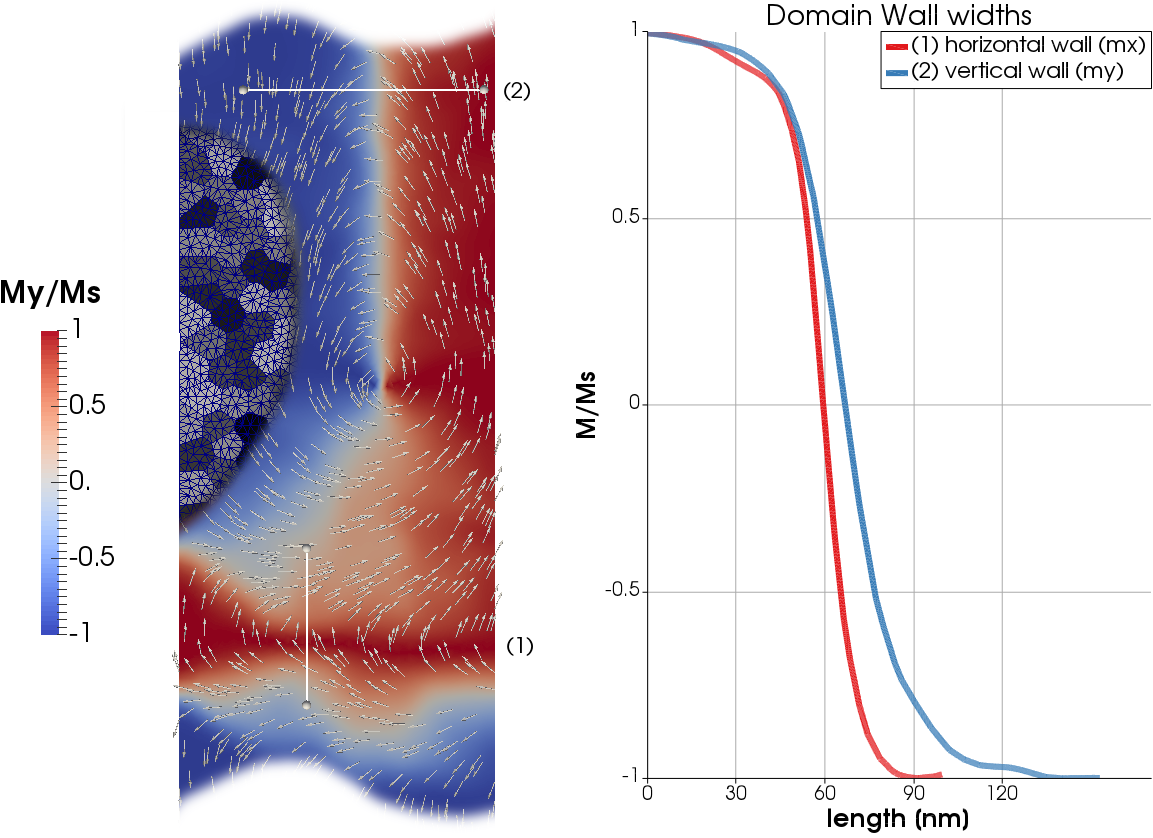}
\caption{Domain walls in 200nm x 1200nm x 5nm thin film structure with an average grain size of $19.5nm$ at $H_{ext,x} = H_c$. Domain walls are much wider than what would be expected by the usual estimates of $l=\pi\sqrt{\frac{A}{K_1}}$ or $l=\pi\sqrt{\frac{\mu_0 A}{J_s^2}}$}
\label{domainwalls}
\end{center}
\end{figure}
For the material parameters used in our simulations ($A=1.5\times 10^{-11}\frac{J}{m}$, $K_1=100\frac{kJ}{m^3}$), the intrinsic exchange length $L_{0,crystalline}$ \eqref{eq:L0} due to the magnetocrystalline anisotropy is $12.25nm$. The maximum in the plots (Fig.~\ref{results}), which is reached when $D = L_{ex}$, as well as the analysis of domain walls (Fig. \ref{domainwalls}) in our simulations suggest that the renormalized exchange length $L_{ex}$ is much larger at around $30nm-40nm$. In addition to exchange softening due to $D<L_{ex}$ we should acknowledge that in thin films, the exchange length is not a well defined quantity at all. For bulk materials the exchange length can be estimated by the classical value of $L_{0} = \sqrt{\frac{A}{K_1}}$ or in the case of soft magnetic materials $L_0 =\sqrt{\frac{\mu_0 A}{J_s^2}}$ with a corresponding estimate for domain wall widths of $l = \pi L_0 = \pi\sqrt{\frac{A}{K_1}}$ and $l=\pi\sqrt{\frac{\mu_0 A}{J_s^2}}$ respectively. Thin films however seem to exhibit a wide variety of possible domain configurations~\cite{schafer2000domains} with different domain wall widths, including instances of 'continuously flowing'~\cite{schafer2001continuous} magnetization configurations where domains in the strict sense are breaking down altogether. This is not inconsistent with the assumption that the random anisotropy model can be applied for thin films, because the anisotropy is still beeing suppressed inside one exchange volume, only the volume is modified. Nonetheless, since the exchange length does not serve as a parameter for the micromagnetic simulations, $L_{ex}$ should naturally arise from the simulations because the coercive fields should reach a maximum when average grain size and exchange length are equal.

\section*{Results}
\begin{figure}[t]
\includegraphics[width=20pc]{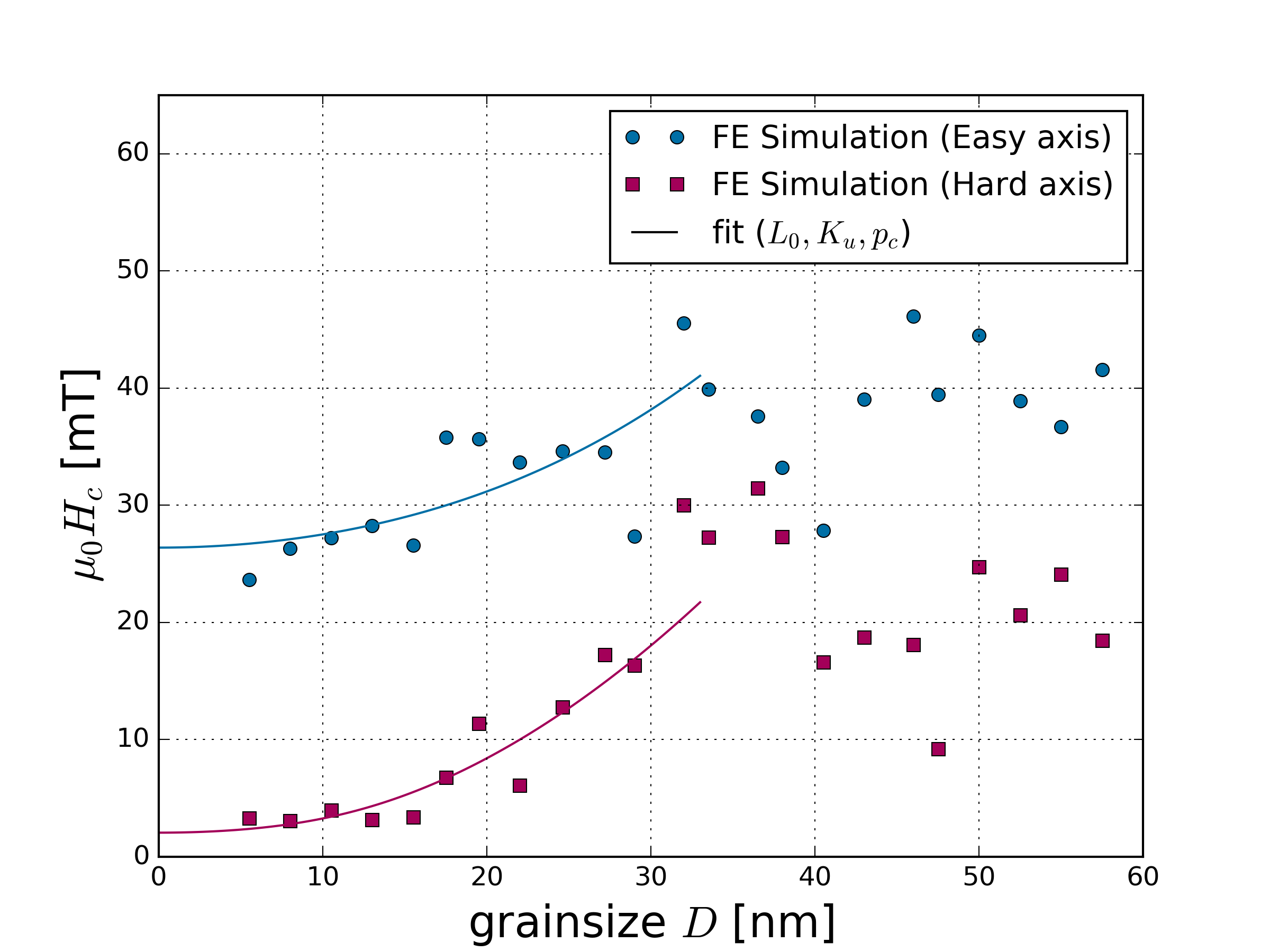}
\caption{Finite elements simulation of easy axis ($\vec{H} = \vec{e}_yH_y$) and hard axis ($\vec{H} = \vec{e}_yH_y$) hysteresis loops for a 200nm x 1200nm x 5nm particle with varying grain size $D$. The numerical results are fitted with a 2D random anisotropy model for materials with additional long range anisotropies~\eqref{eq:RAMHC}. Fit parameters are given in Table~\ref{tbl:fitparams}. \vspace{3pc}}
\label{results}
\end{figure}

\begin{table}
\begin{tabular}{@{}lll@{}}
\toprule
  & easy axis & hard axis \\ \hline
$L_{0,fit}$ 		& $33.94nm$	& $28.76nm$	\\
$p_{c,fit}$ 		& $0.3548$	& $0.2277$	\\
$K_{u,fit}$ 		& $103.6\frac{kJ}{m^3}$ & $12.59\frac{kJ}{m^3}$ \\
$K_{u,amorphous}$  	& $101.4\frac{kJ}{m^3}$ & $20.79\frac{kJ}{m^3}$ \\
\bottomrule
\end{tabular}
\caption{Fit parameters for 2D random anisotropy model with additional long range anisotropy and numerical data from micromagnetic simulations~(Fig. \ref{results}). $K_u(\theta)$ should not be interpreted as one specific (shape) anisotropy. It rather represents an effective anisotropy due to the demagnetization field for the given direction of measurement, since different axes for externally applied fields lead to different magnetization reversal mechanisms (Fig. \ref{fig:hysteresis} \& \ref{fig:magframes}).}
\label{tbl:fitparams}
\end{table}
\begin{figure}[h!]
  \includegraphics[width=20pc]{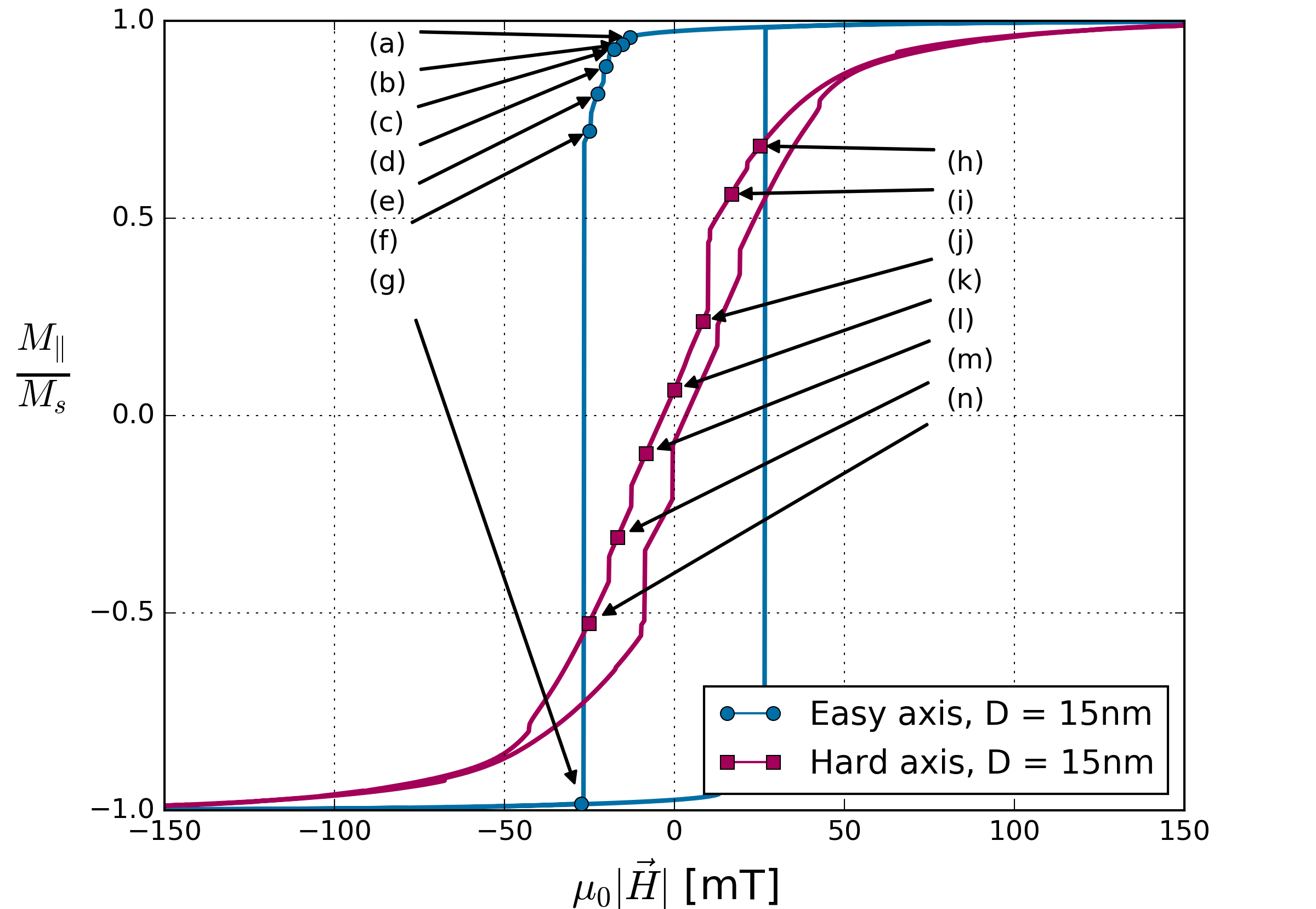}
\caption{Hysteresis loop for easy and hard axis direction for average grain size $D=15nm$. The marked points correspond to magnetization configurations in fig. \ref{fig:magframes}. $M_{\|}$ is the magnetization component parallel to the applied field, meaning $M_y$ for easy axis loops, and $M_x$ for hard axis loops.}
\label{fig:hysteresis}

\vspace{2pc}
  \includegraphics[width=20pc]{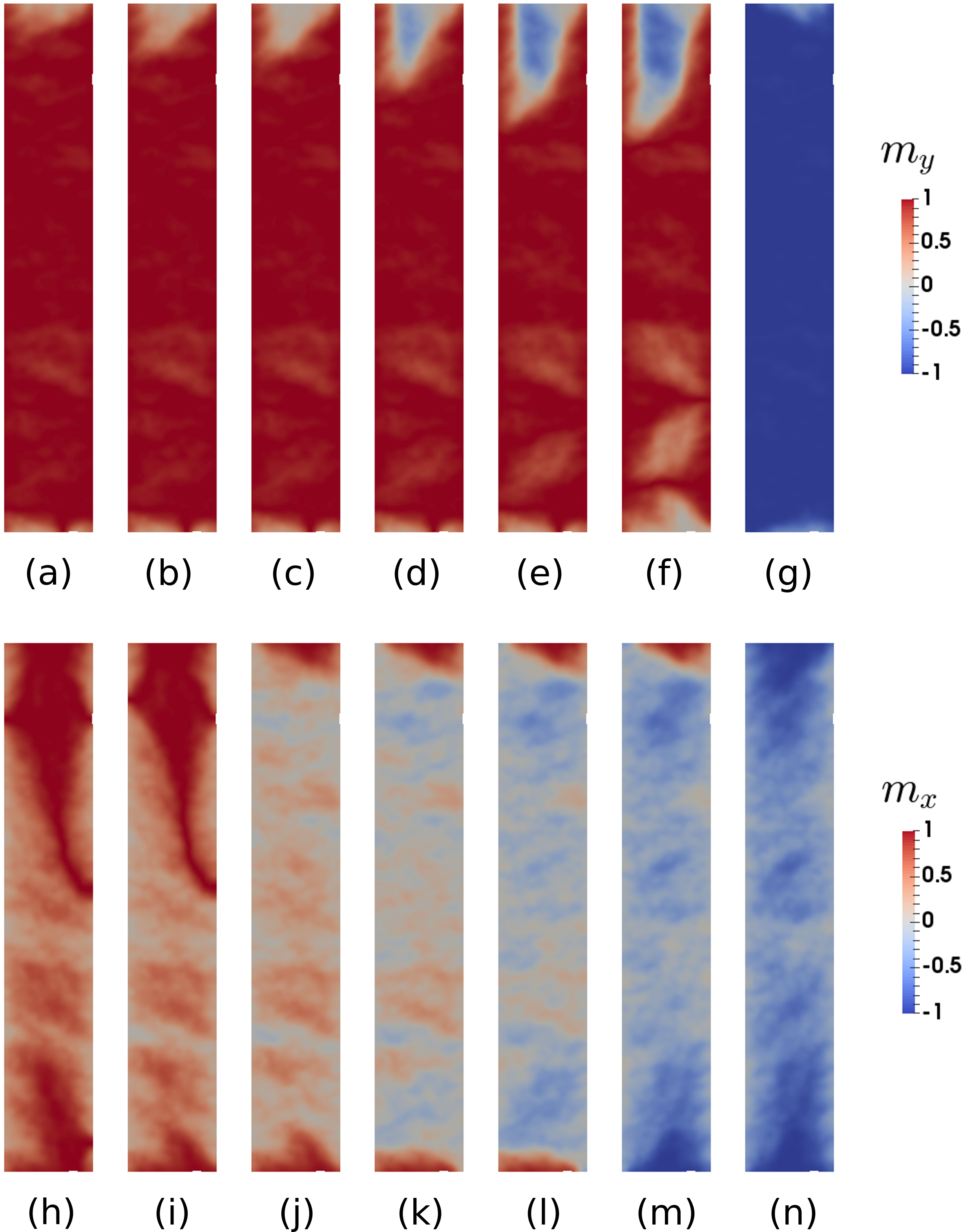}
  \caption{Magnetization configurations for points along the hysteresis loops in fig. \ref{fig:hysteresis}. In easy axis direction (a-g), the magnetization reversal is sudden and simultaneous across the whole particle while in the hard axis direction (h-n), the magnetization reversal is continuous and soft in the majority of the particle with exception of the end-domains which effectively act like a sub-particle with anisotropy in x-direction, leading to a non-zero coercive field for the hard-axis hysteresis loop.}\label{fig:magframes}
\end{figure}

Coercive fields were evaluated for easy and hard axis loops as mean of rising and falling edge coercive fields
\begin{equation}
H_c = \frac{\sum\limits_i^{2N} abs(H_{i,j}\vert_{M_j=0})}{2N}
\end{equation}
where $j$ is $x$ for a hard axis loop, $j$ is $y$ for an easy axis loop and N is the number of simulated field periods. Resulting coercitivities over grain size are shown in Figure~\ref{results}. The simulated values were then fitted with the theoretical model \eqref{eq:RAM2DKuModel} by~\cite{herzer1990grain}
\begin{equation}
\label{eq:HctoK}
H_c = \frac{p_c}{J_s}\langle K \rangle
\end{equation}
where $p_c$ is a dimensionless factor of the order of one. With the full expression for $\langle K \rangle$, \eqref{eq:HctoK} expands to
\begin{equation}
\label{eq:RAMHC}
H_c = \frac{p_c}{J_s} \left[ \frac{K_1}{2} \left(\frac{D}{L_0}\right)^2 + \sqrt{\left(\frac{K_1}{2} \left(\frac{D}{L_0}\right)^2\right)^2 + K_u^2(\theta)} \right]
\end{equation}
where $K_u$ becomes $K_u(\theta)$ because, due to different magnetization reversal modes, $K_u$ manifests differently for different directions of measurement, which is treated in more detail later on in this section.
Fine particle theory predicts $p_c = 0.96$ for uniaxial single crystals, $p_c = 2$ for cubic single crystals and $p_c = 0.64$ for an ensemble of cubic grains with random anisotropy orientation~\cite{bozorth1951ferromagnetism}. $L_0$, $p_c$ and $K_u$ were used as fit parameters (Table~\ref{tbl:fitparams}) while $K_1 = 100\frac{kJ}{m^3}$ and $J_s=1.75T$ were held constant as used in the simulation parameters. $p_c$ fit values of $0.355$ and $0.228$ for easy axis and hard axis respectively are in good agreement with the results of fine particle theory and match very well with experimental data analyzed by Herzer~\cite{herzer1995soft} which yields $p_c\approx 0.2_{\pm 0.1}$. The fit results for $L_0$ of $33.94nm$ and $28.76nm$ are well in the range of exchange lengths observed in simulated domain patterns for 200nm~x~1200nm~x~5nm structures which is covered in more detail in the previous section. Finally, the fitted anisotropy constant $K_u$, which via equation \eqref{eq:HctoK} is related to the y-intersect in Figure~\ref{results}, is compared to simulations with vanishing anisotropy constants ($K_1 = 0\frac{J}{m^3}$) but elsewise unchanged parameters, which corresponds to the amorphous case of vanishing grains $D\rightarrow0$. The resulting anisotopy constants were calculated by \eqref{eq:HctoK} from the coercive fields in the 'amorphous' simulation using the fitted $p_c$ values (Table~\ref{tbl:fitparams}) for the respective direction.
It is not intuitively clear why our structures show one anisotropy constant $K_u(\theta)$ in y- and another anisotropy in x-direction. In fact we do not have one anisotropy with different constants in different directions, but different particle regions each with their own distinct anisotropy. $K_u(\theta)$ represents an effective anisotropy constant for the whole structure and a given direction of measurement. The majority of the particle, which we call the body region, has an anisotropy aligned with the long axis. The top and bottom end regions however have an anisotropy which is aligned with the short edge of the particles, leading to an overall C-,  or S-state like magnetization configuration in the absence of external magnetic fields (see for instance Fig. \ref{fig:magframes}k). In that sense, the descriptions 'easy axis loop' and 'hard axis loop' are technically only correct for the body region of our structure, but since that region makes up the vast majority of the overall volume we shall keep this naming convention. In the case of an easy axis hysteresis loop, the coercive fields are largely unaffected by the anisotropy of the top and bottom regions and effectively only depend on the body anisotropy and the random anisotropy contributions while in the hard axis case, the coercive fields are in turn unaffected by the body anisotropy and only depend on the top and bottom anisotropies and the random anisotropy contributions because a hysteresis loop perpendicular to a uniaxial anisotropy axis does not induce a coercive field.\\
Looking, for instance, at the hysteresis loops of our $D=15nm$ simulations (fig. \ref{fig:hysteresis}), we can see a very stoner-wohlfarth-like easy axis loop with a sudden magnetization reversal that is almost simultaneous across the whole particle. The hard axis loop also behaves Stoner-Wohlfarth-like for the most part, with exception of the top and bottom regions of the particle, where so called end-domains persist (C-state). At zero external field, the inner region of the particle has an average magnetization x-component of zero, but the average magnetization of the end-domains is still oriented in +x direction (figs. \ref{fig:hysteresis} and \ref{fig:magframes}k), leading to a non-zero coercive field. This example shows that even a particle with a shape as simple as a rectangle, can display complex manetization dynamics.

\section*{conclusion}
Hysteresis loops of 200nm x 1200nm x 5nm CoFe particles with nanocrystalline grains ranging from 5nm to 60nm were simulated using the finite elements package {\it femme}~\cite{suess2002time}. Coercive fields in easy-axis as well as in hard-axis direction were evaluated as a function of grain size D to investigate the interplay of shape anisotropy and random crystalline anisotropy. The results were then fitted with a Random Anisotropy Model for two dimensions which has been extended for additional long range anisotropies to account for the shape anisotropy induced by the demagnetizing field. Our results show that coercivity in nanocrystalline thin films is well described by the 2D extended Random Anisotropy Model~\eqref{eq:RAM2DKuModel} for both easy, and hard direction. The interplay between random crystalline and uniform anisotropy which is not easily accessible theoretically for the general 3-dimensional case, has been illustrated for the 2-dimensional case of thin-films and well reproduced by our micromagnetic simulations. \\
Although there have been efforts to describe the grain size regime of $D>L_{ex}$~\cite{herzer1990grain,herzer1993nanocrystalline,herzer1995soft} for bulk materials, this work does not delve into the process of preparing samples with significantly larger grain diameters than film thickness. Furthermore the structures investigated in this work were to small to accomodate a large enough number of grains to provide statistically significant results at large grain sizes. The authors plan to perform corresponding experiments with larger structures in the future, as right now the computational capacities are insufficient.\\

\section*{Acknowledgements}
The financial support by the Austrian Federal Ministry
of Science, Research and Economy, the National Foundation
for Research, Technology and Development and the Austrian
Science Fund (FWF): F4112 SFB ViCoM, is gratefully acknowledged.

\bibliography{references}
\end{document}